# Large Earthquake Prediction Methods on the Base of INFREP VLF / LF Data

Manana Kachakhidze[1], Nino Kachakhidze-Murphy[1], Badri Khvitia[2], Giorgi Ramishvili[3]

[1] Georgian Technical University, 77 Merab Kostava St, T'bilisi 0175, Georgia
[2]Ivane Javakhishvili Tbilisi State University, 1 Chavchavadze Avenue, Tbilisi 0179, Georgia
[3]Ilia State University, 3/5, Kakuca Cholokashvili Ave, Tbilisi 0160, Georgia

Corresponding author Manana Kachakhidze: kachakhidzem@gmail.com

## Abstract

In the presented paper the possible methods of the large earthquake prediction are offered. During the study, it was used data of the INFREP (European Network of Electromagnetic Radiation) existent before earthquake.

The elaborated methods given in the work are capable of simultaneous determination of all three parameters necessary for large earthquake prediction (magnitude, epicenter and time of occurring) with certain accuracy.

## 1. Introduction

Studies of earthquake problems in the world were especially intensified from the second half of the last century, since alongside with theoretical studies it became possible to carry out high level laboratory and satellite experiments (Biagi et al., 1999; Hattori et al., 2004; Parrot, 2006; Freund et al., 2006; Varotsos, et al., 2006; Bleier et al., 2009; Eftaxias et al., 2009, Contadakis et al, 2010; Ouzounov et al., 2011; Biagi et al., 2013; Tramutoli et al.,2013).Thanks to them in the earthquake preparation process various anomalous changes of geophysical fields have been revealed in lithosphere as well as in atmosphere and ionosphere.

In recent decades in some seismic active countries of the world the network for detection of VLF/LF radiosignals have been organized.

The role of abovementioned electromagnetic emissions networks is very important because it enriched science with invaluable information and made the searching of earthquake forecasting problem far more wide-scale and integral.

Our research team has created a work (Kachakhidze et al., 2015), where the frequency of electromagnetic radiation existing in the earthquake preparation period analytically is connected with the fault length of incoming earthquake (1)

$$l = \beta \frac{c}{\omega} \qquad (1)$$

where ω is the frequency of existent electromagnetic emissions, c is light speed, β is the characteristic coefficient of geological medium (it approximately equals to 1).

Based on this work, the theoretical method of earthquake prediction was developed (Kachakhidze et al., 2016). The relevant results wer reported at the EGU Assembly in 2016 (Kachakhidze et al., 2016 A).

In this presented work we have tried to check earthquake prediction possibilities on the base of the INFREP retrospective data and abovementioned papers. At this stage of study we do not discuss foreshocks and aftershocks series.

As noted above, the theory created by our group (Kachakhidze et al., 2016) is based on the analysis of frequencies of electromagnetic radiation existing in the earthquake preparation period.

Because INFREP network fixes the every minute amplitudes of 10 different frequency electromagnetic radiation, based on INFREP data, it was calculated and used the every minute frequency numerical values transformed by the normal distribution of Gauss (Jenkins et al., 1972).

## 2. Disscussion

### 2.1. Determination of the Incoming Earthquake Magnitude

As it is known, earthquake magnitude is measured based on seismic waves data and only after the earthquake, because, before earthquake we do not have information about parameters required to determine the magnitude.

In this regard, electromagnetic radiation has different feature: it arises during the earthquake preparation period, exists until the last aftershock, and more importantly, we can fix it as soon as it originates.

This is the main advantage of the electromagnetic emissions because before the earthquake, in advance, by EM emissions records, it is possible to determine the length of the fault in the focus (1) (Kachakhidze et al., 2016) as well as magnitude (2, 3) of the incoming earthquake:

$$lgl = 0.6\ M_s - 2.5 \qquad (2)$$

$$M_w = 4.38 + 1.49 * \log l \qquad (3)$$

According to the baseline frequencies of the INFREP receivers, by the abovementioned (1), (2) and (3) formulas, we calculated the fault lengths and magnitudes. In order easily to mention the frequency channels in this article, they are symbolically marked with letters (Table 1).

Table I. INFREP frequency channels with relevant fault length and magnitudes

| frequency | channels | fault | Ms | Mw |
|---|---|---|---|---|
| (Hz) | | length in meters | | |
| 20 270 | C | 14 800 | 6.1 | 6.1 |
| 20 900 | D | 14 354 | 6.1 | 6.1 |
| 23 400 | E | 12 820 | 6.0 | 6.0 |
| 37 500 | F | 8 000 | 5.7 | 5.7 |
| 45 900 | G | 6 535 | 5.5 | 5.6 |
| 153 000 | H | 1 960 | 4.7 | 4.8 |
| 162 000 | I | 1 851 | 4.6 | 4.8 |
| 183 000 | J | 1 639 | 4.5 | 4.7 |
| 216 000 | K | 1 388 | 4.4 | 4.6 |
| 270 000 | L | 1 111 | 4.2 | 4.4 |

The following classifications were made: channels that indicate the M≥5 earthquakes corresponding frequencies (C, D, E, F, G) are considered as “strong”channels, and "weak" channels, which frequencies correspond to 4.2<M<5.5 earthquakes (H, I, J, K, L) (Table 1).

The calculations show that the frequencies are highly sensitive to the fault length. Therefore, the frequency of electromagnetic radiation allows us to measure an earthquake magnitude much more precisely than it measures today. Seismologists and focus physicists should decide in the future whether it is necessary to determine magnitude with such high accuracy.

### 2.2. Separation of the Active Channel

In order to possible predictability of the large earthquakes, studies have been conducted for Crete earthquake with M= 5.6 (25/05 / 2016, 08:36:13 UTC) for period 04.04. 00: 00-16.06.23: 59: 00 (73 days).

As known, avalanche – unstable process of fault formation is important stage in the earthquake preparation period (Mjachkin et al., 1975). The avalache - unstable process is exactly the process during which the significant changes of geophysical fields take place in the focus. Because of IAI (lithosphere-atmosphere-ionosphere) coupling system, these changes reveal themselves as earthquake precursors or indicators on the earth surface, atmosphere and ionosphere.

Duration of avalanche process of fault formation mainly varies from 10-14 days to 1 month before earthquake and it depends on geological peculiarities of the region (Mjachkin et al., 1975). According to our theory (Kachakhidze et al., 2016, 2016 A), EM radiation is the main precursor of earthquake. Therefore, the earthquake preparation process should be reflected in the INFREP records.

Because we searched the Crete earthquake based on retrospective data, we had an opportunity to study the full earthquake preparation picture during the reviewed period.

For considered period, amplitudes and frequencis graphs have been created separately for receiver all channels. As expected, graphs for these two parameters unequivocally are identical, because we represent the amplitude graph for two strong channels: F (37 500 Hz) and C (20 270 Hz) (Fig.1, Fig. 2. Earthquake occurring moment is noted by arrow).

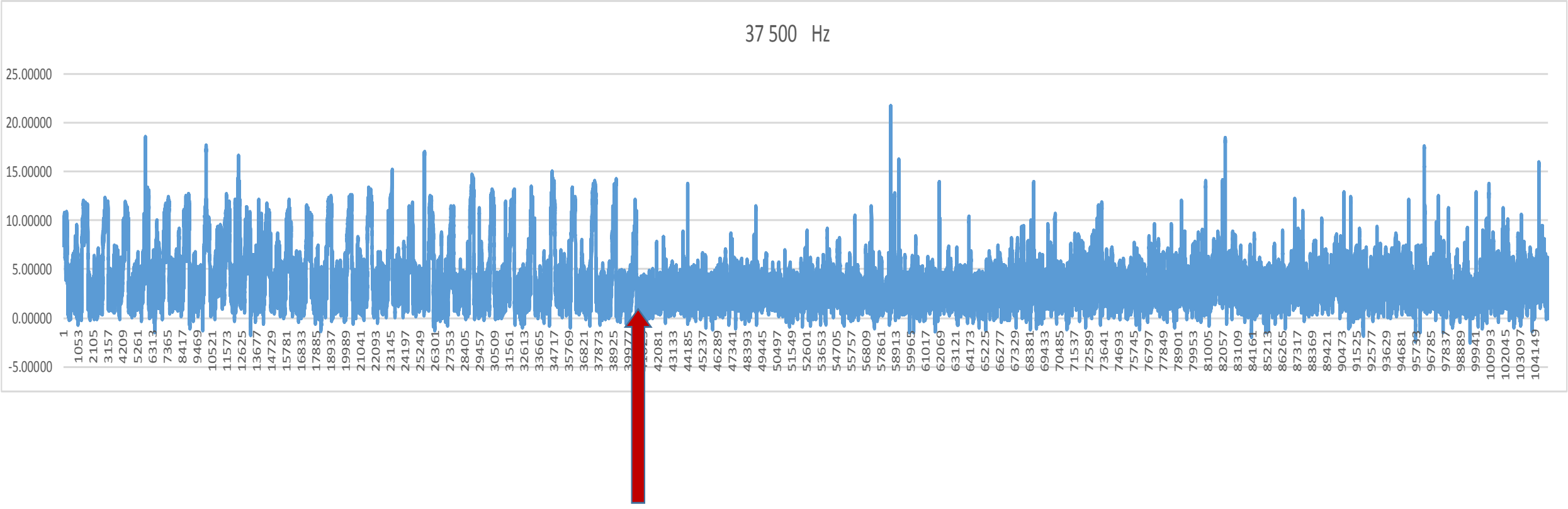

Fig. 1. 73-days amplitude graphs for F channel

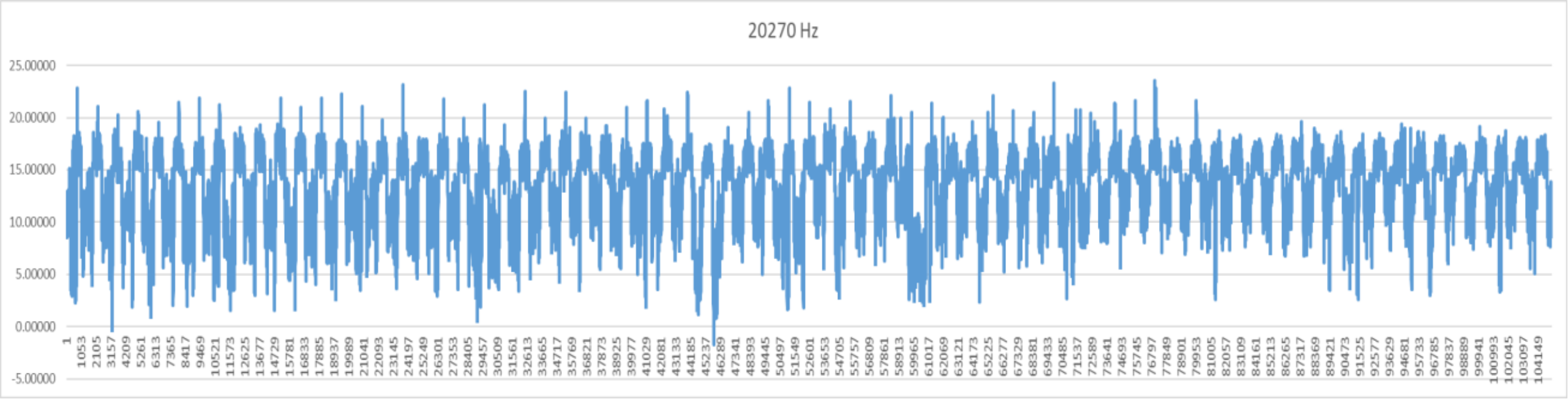

Fig. 2. 73-days amplitude graphs for C channel

By discussion of the graphs we identified their general characteristic feature - on all 10 channels, except F (37 500 Hz), during the reviewed period, diurnal periodic variations are clearly expresed. We have such variations in the F channel recordings too, but till some period, in particular, to 02.05, after which the anomalous process of starts, indicating that the avalanche – unstable process of fault formation already began in earthquake preparation period.

In order to make the ongoing processes more visible, additionally we created graphs of several days by all 10 channels data. The 3 and 4 figures show the graps of exactly 8-days data which consist with the beginning moment of avalanche –unstable process on the same 37 500 Hz and 20 270 Hz channels.

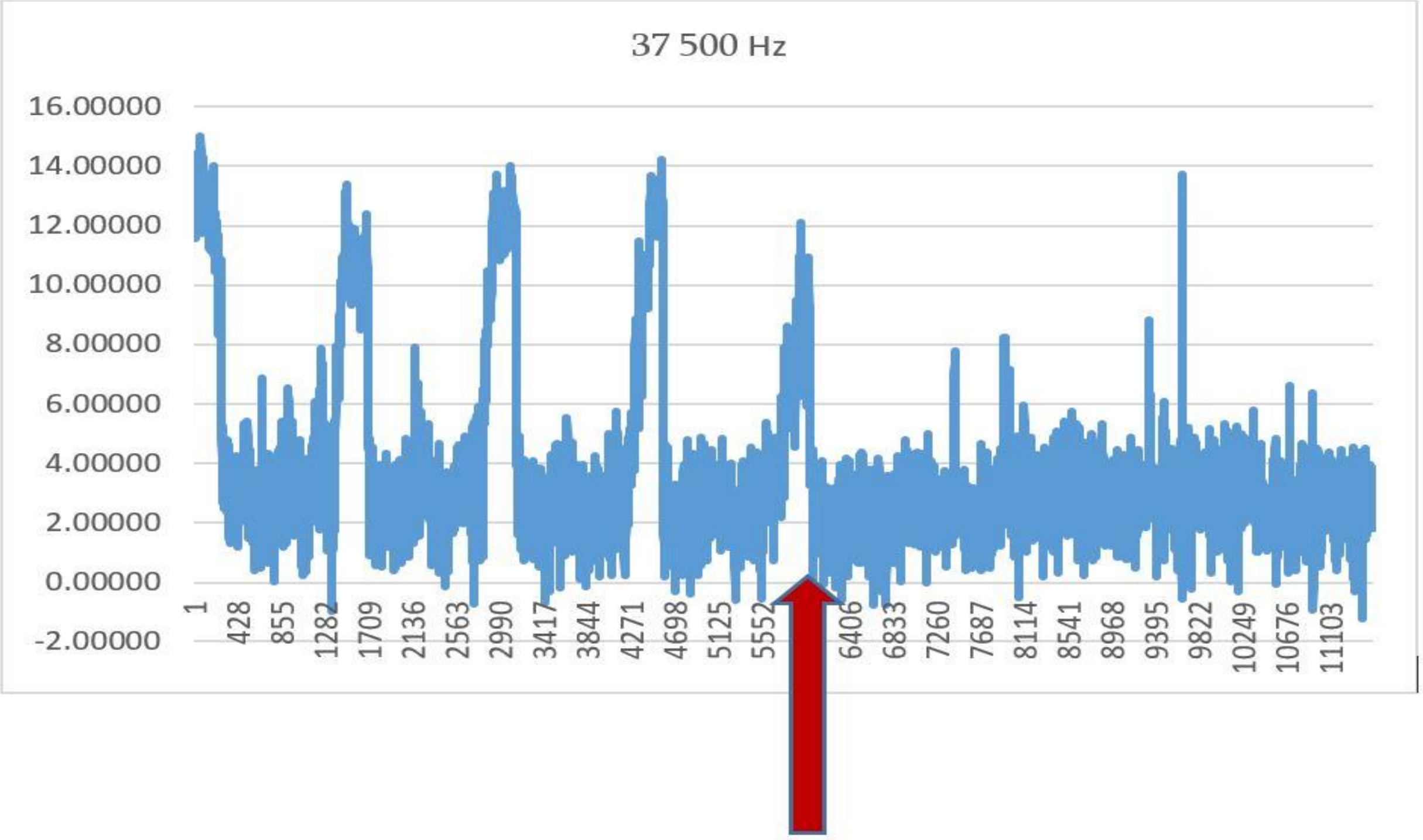

Fig. 3. F channel amplitude graphs for 28 April- 5 May

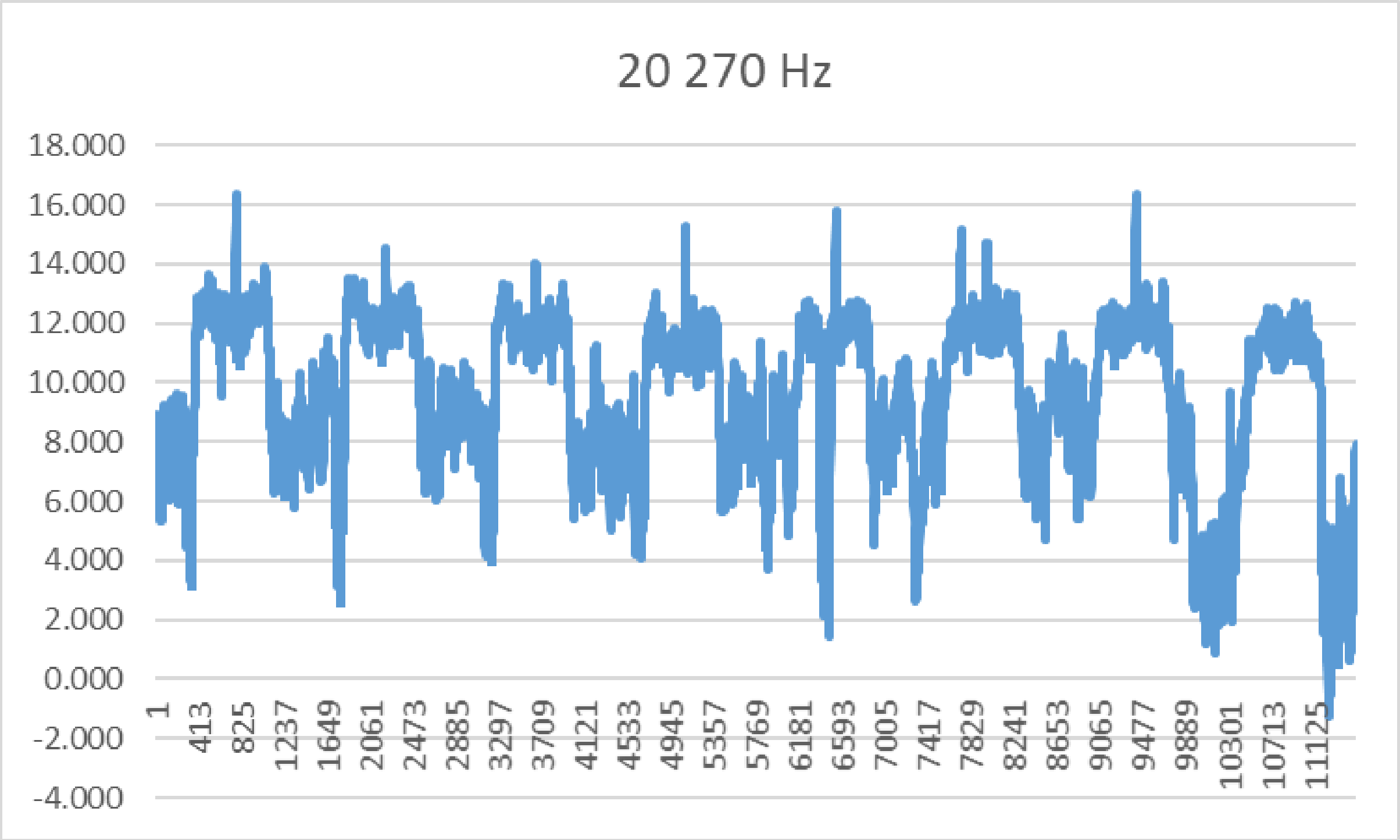

Fig. 4 . C channel amplitude graphs for 28 April- 5 May

The general analysis of 8-days graphs has shown that during the large earthquake preparation period, not every channel, among them strong, can "see" the earthquake preparation process. It should be assumed that this process is reflected in the channel, in our case F (37 500 Hz) "active" channel, on which the stage of avalanche process of fault formation is expresed and based on data of which should be made predictable conclusions in the future.

Because our goal is to find large earthquake forecasting methods (on example of Grete earthquake), we used only so-called “strong” channels (C-G, Table I) data.

As shown above, the F (37 500 Hz) "active" channel was detected by analyzing the full 73-days retrospective data. However, in case of the earthquake monitoring process, when the conclusions should be made in advance, before the earthquake, it is obvious that such an approach will not work.

Let us recall that in relatively early stages of the earthquake preparation, the origin and locking of small chaotic orientation cracks take place. Later, since the beginning of avalanche process of fault formation, the cracks start oriented locating in parallel planes to each other on so-called “cracked strip", which is followed by their joining and formation of main fault at the last stage of earthquake preparation (Mjachkin et al., 1975).

If any frequency channel actually reflects the earthquake preparation, it is natural that the relevant geological process should be reflected in the frequency data since there is an analytical connection between the frequency of EM radiation and the fault length in the earthquake focus (Kachakhidze et al., 2015).

For this reason, we separately calculated the total length of the every minute cracks corresponding to C-G frequency channels towards the length relevat to channel baseline frequency. After we computed the average daily values of this signficances. The results of the calculation are given in Table II.

Table II. Average daily values of cracks total length in percentages for C-G channels

| Data | 20270 Hz (C) | 20 900 Hz (D) | 23 400 Hz (E) | 37 500 Hz (F) | 45 900 Hz(G) |
|---|---|---|---|---|---|
| 4 April | 72,9 | 72,3 | 73,7 | 93,7 | 87,2 |
| 5 | 71,7 | 71,8 | 73,6 | 93,8 | 86,9 |
| 6 | 71,8 | 70,8 | 73,4 | 93,3 | 88,2 |
| 7 | 71,7 | 74 | 73,4 | 94,1 | 86,9 |
| 8 | 72,9 | 73,2 | 73,3 | 93,5 | 87,4 |
| 9 | 72,5 | 71,8 | 73,3 | 93,3 | 87 |
| 10 | 72,2 | 71,9 | 73,4 | 93,5 | 87,3 |
| 11 | 72,8 | 73,1 | 73,7 | 92,7 | 87,1 |
| 12 | 72,6 | 80,3 | 73,9 | 92,7 | 87,5 |
| 13 | 72 | 78,8 | 75,1 | 93,3 | 87,9 |
| 14 | 71,8 | 80,3 | 75,9 | 93,6 | 87 |
| 15 | 72,2 | 79,2 | 74,5 | 93,7 | 87,4 |
| 16 | 71,8 | 70,1 | 74,3 | 94 | 87,2 |
| 17 | 71,8 | 72 | 74,6 | 93,8 | 87,5 |
| 18 | 72,3 | 72,1 | 76,2 | 93,5 | 87,4 |
| 19 | 71,4 | 71,9 | 76,8 | 93,5 | 87,2 |
| 20 | 71,8 | 71,7 | 76,7 | 93,2 | 88,3 |
| 21 | 71,4 | 72,1 | 74,4 | 94,2 | 87,2 |
| 22 | 71,9 | 72,2 | 73,4 | 94,2 | 87,5 |
| 23 | 71,8 | 72,1 | 73,5 | 94,8 | 87,2 |
| 24 | 73,7 | 72,3 | 73,9 | 94,1 | 87,6 |
| 25 | 72,8 | 72,2 | 73,7 | 94 | 87,6 |
| 26 | 72,2 | 71,8 | 73,7 | 94,1 | 87,7 |
| 27 | 71,1 | 74,4 | 73,7 | 94,3 | 88,6 |
| 28 | 71,2 | 72 | 73,6 | 93,8 | 88 |
| 29 | 70,8 | 72 | 73,3 | 94,2 | 87,2 |
| 30 | 71,2 | 72 | 73,4 | 94,3 | 87,4 |
| 1 May | 71,6 | 72,1 | 73,5 | 94,5 | 88,1 |
| 2 | 71,8 | 73,4 | 73,7 | 95,2 | 88,1 |
| 3 | 71,2 | 71,6 | 73,5 | 95,9 | 87,7 |
| 4 | 71,8 | 73,2 | 75 | 95,9 | 90 |
| 5 | 74,9 | 76,5 | 78,4 | 96,1 | 89 |
| 6 | 71,5 | 72,7 | 74 | 95,8 | 87,5 |
| 7 | 71,1 | 72,3 | 73,8 | 95,9 | 87,1 |
| 8 | 71,3 | 72,3 | 73,8 | 96,2 | 87,3 |
| 9 | 72,8 | 72,3 | 75,8 | 96,1 | 88,1 |
| 10 | 71,1 | 71,9 | 77,5 | 95,9 | 87,5 |
| 11 | 71,1 | 72,8 | 77,2 | 95,6 | 87,4 |
| 12 | 70,7 | 71,8 | 76,4 | 95,7 | 87,3 |
| 13 | 70,9 | 73,5 | 77,2 | 95,9 | 88,1 |
| 14 | 70,6 | 72,2 | 77 | 95,8 | 87,7 |
| 15 | 74,1 | 76,1 | 81,2 | 95,6 | 89,1 |
| 16 | 73,1 | 72,4 | 79,9 | 95,6 | 89 |
| 17 | 71,4 | 68,4 | 76,6 | 95,5 | 88 |
| 18 | 71,2 | 71,8 | 76,8 | 95,8 | 89,3 |
| 19 | 71,2 | 72,5 | 77,2 | 95,4 | 88,3 |
| 20 | 71,6 | 72,8 | 77 | 95 | 88,2 |
| 21 | 70,9 | 72,1 | 76,9 | 95,1 | 88,3 |
| 22 | 70,9 | 73,1 | 77,3 | 95,2 | 88 |
| 23 | 72,1 | 72,3 | 77,6 | 94,7 | 88,3 |
| 24 | 71 | 72 | 76,9 | 93,8 | 88,5 |
| 25 | 70,9 | 74,9 | 76,7 | 95,2 | 89,6 |
| 26 | 70,5 | 69,8 | 77,1 | 95,3 | 88,4 |
| 27 | 70,4 | 72 | 76,9 | 94,8 | 88 |
| 28 | 70,9 | 73,3 | 76,9 | 94,5 | 88,5 |
| 29 | 70,6 | 71,9 | 77,2 | 94,5 | 88,1 |
| 30 | 71,8 | 71,7 | 77,8 | 93,8 | 88,4 |
| 31 | 70,8 | 71,9 | 77,8 | 93,9 | 88,1 |
| 1 June | 70,5 | 71,8 | 76,9 | 95,2 | 89,5 |
| 2 | 70,8 | 71,9 | 77 | 95,4 | 88,4 |
| 3 | 70,8 | 71,9 | 77,2 | 95 | 88 |
| 4 | 70,8 | 71,8 | 76,5 | 95,1 | 88,2 |
| 5 | 71,1 | 71,9 | 76,8 | 94,6 | 88,1 |
| 6 | 72,2 | 71,8 | 76,8 | 94,7 | 88,3 |
| 7 | 70,6 | 71,8 | 76,8 | 94,6 | 88,3 |
| 8 | 70,8 | 72,1 | 77,1 | 94,5 | 89,3 |
| 9 | 71,1 | 72 | 76,9 | 94,7 | 88,8 |
| 10 | 71,3 | 71,8 | 76,7 | 94,7 | 88,8 |
| 11 | 70,7 | 71,8 | 77,1 | 95,7 | 88 |
| 12 | 70,8 | 72,2 | 77,2 | 95,5 | 88 |
| 13 | 71,6 | 71,7 | 77,1 | 94,3 | 88,1 |
| 14 | 70,5 | 71,7 | 77,2 | 94,5 | 88,4 |
| 15 | 70,9 | 71,5 | 77,5 | 94,5 | 89,4 |

In order to make the results more visible, the same results are shown in Fig.5.

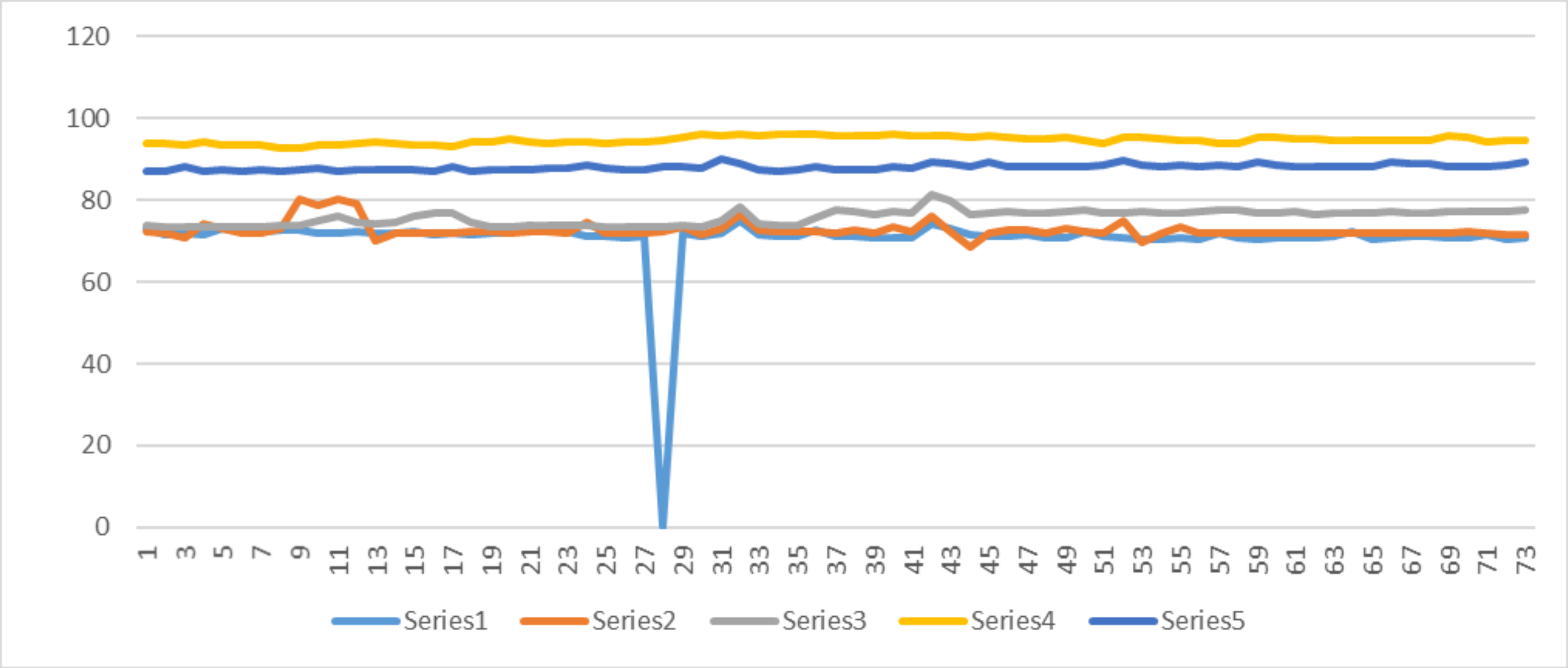

Fig. 5. Average daily values of cracks total length in percentages for C-G channels

It turned out that the F (37 500 Hz) channel is most active and the G (45 900 Hz) channel is less active to the earthquake preparation process, as average daily value of the total length of the cracks (in percentages) was the maximum for these two channels. This means that from disscused 5 channels, only two, with above mentioned frequancies, described the earthquake preparation process. In this case, according to Table I, the magnitude of incoming earthquake should be between 5.5 and 5.7 (Crete earthquake magnitude is really estimated as M= 5.6).

Thus, in case of Crete earthquake about 50 days before the earthquake it is possible to determine the active "cracked strip" on which the main fault is formed in the future and the earthquake occurs. That method of revealing of the active channel will be appropriate for the earthquake preparation process monitoring because such advantage of cracks total length is appeared just since the first records.

During monitoring, step-by-step, there also is possibility to check the obtained result once again, on the active channel, before the earthquake, the avalanche process of fault formaion described by geologocal model, should appear in frequency data (Mjachkin et al., 1975).

Since, in case of discussed earthquake, only F (37 500 Hz) frequency channel meets both conditions: for this channel the average daily value of the total length of the cracks is maximal by per cent and a avalanche process of fault formation appears only on F (37 500 Hz) channel, obviously, to predict the earthquake, we must rely only on the data of this channel.

It is known that the logical end of avalanche process of fault formation is the final stage of earthquake preparation when, at the expense of cracks joining in the focus of incoming earthquke, the main fault forms. This means that (in the case of discussed earthquake), it is expected to form the main fault with the length appropriate to F (37 500 Hz) frequency.

By the formula (1), expected length of the main fault, should be 8 000 meters, but in case of monitoring, it is possible to adjust this numerical value according to frequency significances.

As above said, knowledge of the length of the main fault allow us to determine incoming earthquake magnitude in advance, approximaltely 50 days before earthquake by (2) and (3) formulas.

### 2.3. Possibilities of Determination of the Preparation Area and Epicenter of Incoming Earthquake

Preliminary assessment of the magnitude, approximately 50 days before incoming earthquake (on example of Crete earthquake), gives us an opportunity to quantify the preparation area of incoming earthquake by (4) formula (Dobrovolsky et al., 1979):

$$R \approx 10^{0.43\,M} \quad (4)$$

where R is measured in km.

In addition, epicentral area must has a positive potential (Kachakhidze et al., 2015).Obviously, bad weather or any technogenic process can be the reason for changing of the potential sign of the earth's local area, but it is not a problem to filter the appropriate field. As soon as the active frequency channel is detected, from the points selected around the receiver, where EM emissions are fixed, by the Direction-finding method, it is possible to define the incoming earthquake epicenter.

In order to specify the location of epicenter, we must take into account the data of magnetic and telluric fields, quality of changes of medium geoelectric heterogeneity, TEC anomaly and other parameters that reveal themselves during earthquake preparation period (Hattori, et al, 2004; Moroz et al., 2004; Bleier, et al., 2009; Moldovan et al., 2012; Dudkin, et al.,2013; Biagi et al., 2013; Hayakawa et al.,2013; Kachakhidze et al., 2015 2(N 14); Ouzounov et al., 2018). It does not excluded to determine the direction of fault in advance. However, these issues should be worked out and agreed with earthquake focus physics experts.

Thus, it turned out that revealing of active channel is important for making of prognostic conclusions about not only incoming earthquake magnitude (see 2.2) but epicenter too.

### 2.4. Possibilities of Determination of Incoming Earthquake Occurring Time

As above said, in case of Crete earthquake, only the F (37 500 Hz) active channel reflects the earthquake preparation process. Therefore, through the data of this channel we tried to determine the third parameter, the incoming earthquake time of occurring required for earthquake prediction.

It must note in advance, that from the EM emissions records it can be easily excluded VLF radiation of cosmic origination. In the case of considered earthquake, the "cracked strip", about 50 days prior to earthquake, continuously radiates the frequency approximately equal to frequance at the earthquake occuring moment. As known, the magnetosphere VLF radiation exists in perturbed geomagnetic conditions but these perturbations have no continuous character for 50 days. Therefore, it is easy to separate VLF radiation of the cosmic origin from EM radiation caused by earthquake preparation process.

If any frequency graph shows that the anomalous picture lasts for more than a week, it is possible to assume that it already takes place avalanche-unstable process of fault formation.

According to the Fig.1, fault formation avalanche process of earthquake preparation started 23 days before the earthquake.

As it was noted above, by geological model, the duration of the avalanche process of fault formation depends on the peculiarities of the region, but in general, it lasts from 10-14 days to 1 month. In the case of monitoring, appearing of this process on frequency records, gives us an opportunity to determine the probable time of occurring of incoming earthquake.

The study shows that detailed processing of the data in order to earthquake prediction is reasonable only from the starting moment of avalanche-unstable process.

Because avalanche-unstable process of earthquake preparation means formation of main fault length at the expense of cracks opening and locking in the focus of incoming earthquke, we continued searching by formula (1) to determine character of fault length change in the focus.

Here and thereafter, the data is processed by using average square deviations:

$$\bar{x} = \sum x_i \omega_i \qquad (5)$$

where $\bar{x}$ is average numerical values and $\omega_i$ - $x_i$ - relative frequencies of data.

$$s^2 = \sum \omega_i [(x_i)^2 - (\bar{x})^2] \qquad (6)$$

$$\sigma = \sqrt{s^2} \qquad (7)$$

s and $\sigma$ are dispersion and average square deviation respectively.

Based on the above formulas, we calculated $\bar{x} \pm \sigma$, $\bar{x} \pm 2\sigma$ and $\bar{x} \pm 3\sigma$ values by every minute data of fault length in focus.

Fig. 6 shows results of $\bar{x} \pm 3\sigma$ calculation. It describes process of main fault formation in the focus well enough.

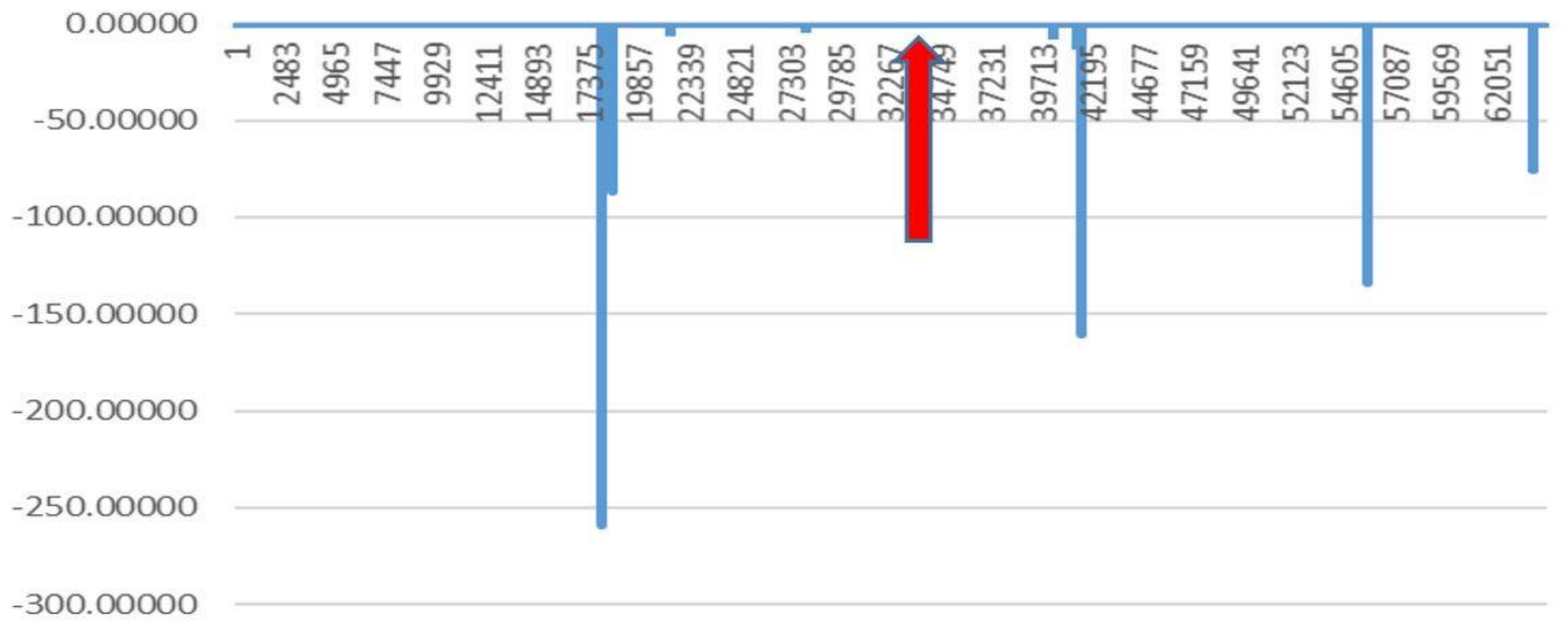

Fig. 6. Fault length changes before Grete earthquake calculated by $\bar{x} \pm 3\sigma$ value

Data analysis shows that 9-10 days prior to the earthquake (on 14-15 May), it happened two main cracks locking on 8-km "cracked strip" in the 9-hour interval. We may presume that in results of it the main part of magistral (final) fault forms.

It is possible that such development of avalanche-unstable process of fault formation in time is characteristic for such magnitude earthquakes of this region. This assumption indicates the necessity of working out the retrospective data for each region in future.

In order to improve about the 20-days forecasting method for considered earthquake, we averaged data of frequencies by 1440 minutes (1 day) and calculated relevant $\bar{x}$ and $\bar{x} \pm \sigma$ values for these data. In this case, just as before, we discussed the only period that involves the avalance-unstable process (Fig. 7).

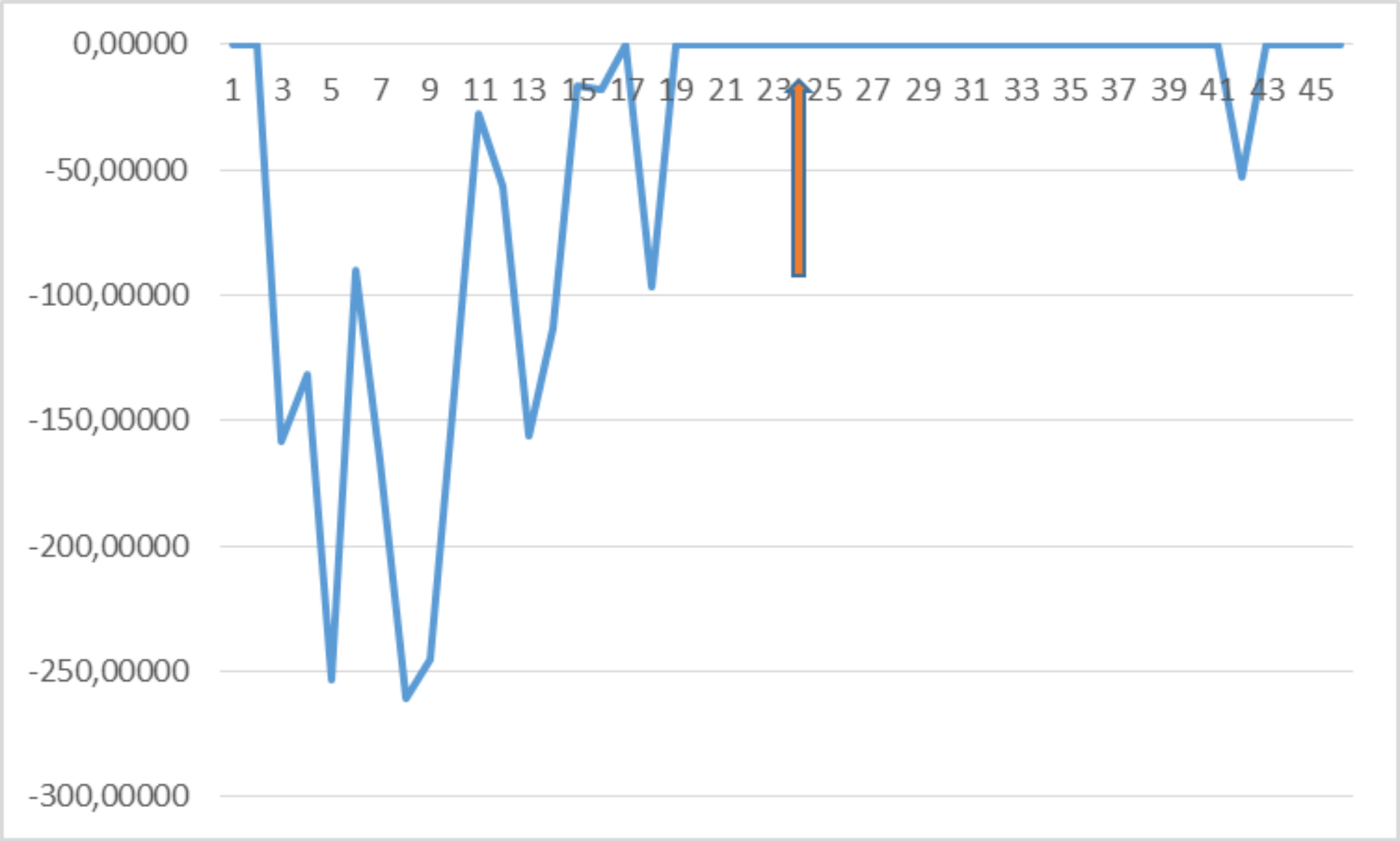

Fig. 7. 19-days avalanche process of fault formation and 5-days EM emissions "silence" period before Grete earthquake calculated by frequency $\bar{x} \pm 3\sigma$ value

Fig.7, similarly to Fig.1, shows that avalanche-unstable process starts 23 days before the earthquake and lasts about for 19 days, which is expressed by sharp changing of frequency values (fault length in the focus).

Such changes of frequencies should be followed by a EM silence period, when almost it is not possible to fix radiation, or the main fault is already formed, on which the certain "portion' of tectonic stress is spent. Of course, the certain time is necessary for restoring of this stress "portion" (Johnston, 1997; Biagi et al.,2009; Boudjada et al., 2010; Papadopoulos et al.,2010; Eftaxias et al. 2013; Kachakhidze et al., 2015; Kachakhidze et al., 2016, 2016 (A)). As soon as the tectonic stress exceeds the limit of rocks strength, the earthquake occurs.

This process is clearly expressed in Fig. 7, where indeed, after the avalanche-unstable process weakening, 5-days EM "silence" period appears, after which earthquake occurs.

At the next stage of the research, we tried to determine incoming earthquake occurring time with much higher accuracy. For this goal, we used the same frequency data avaraged by 1440 minutes (1 day). After that we calculated the frequency ratio and their $\bar{x} \pm \sigma$, $\bar{x} \pm 2\sigma$ and $\bar{x} \pm 3\sigma$ values. The relevant graphs for $\bar{x} \pm \sigma$ and $\bar{x} \pm 3\sigma$ are given on the Figures 8 and 9 correspondingly. The result of the research shows a clear, sharp anomalous changing of frequency 2-days before earthquake, which could be considered as short-term, 2-days prediction of considered earthquake.

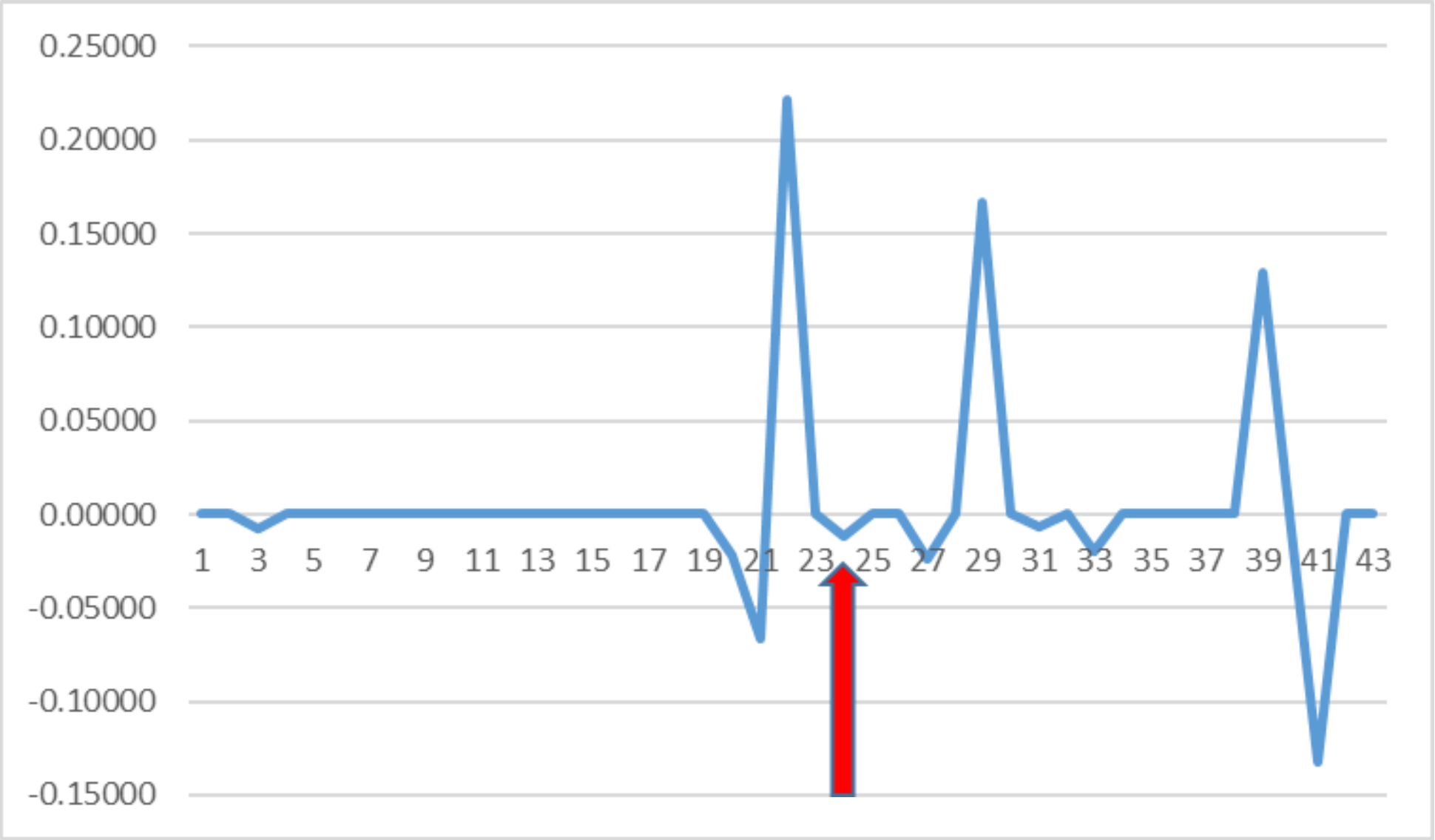

Fig .8. EM emissions “silence” period and anomaly, calculated by frequency ratio $\bar{x} \pm \sigma$ values

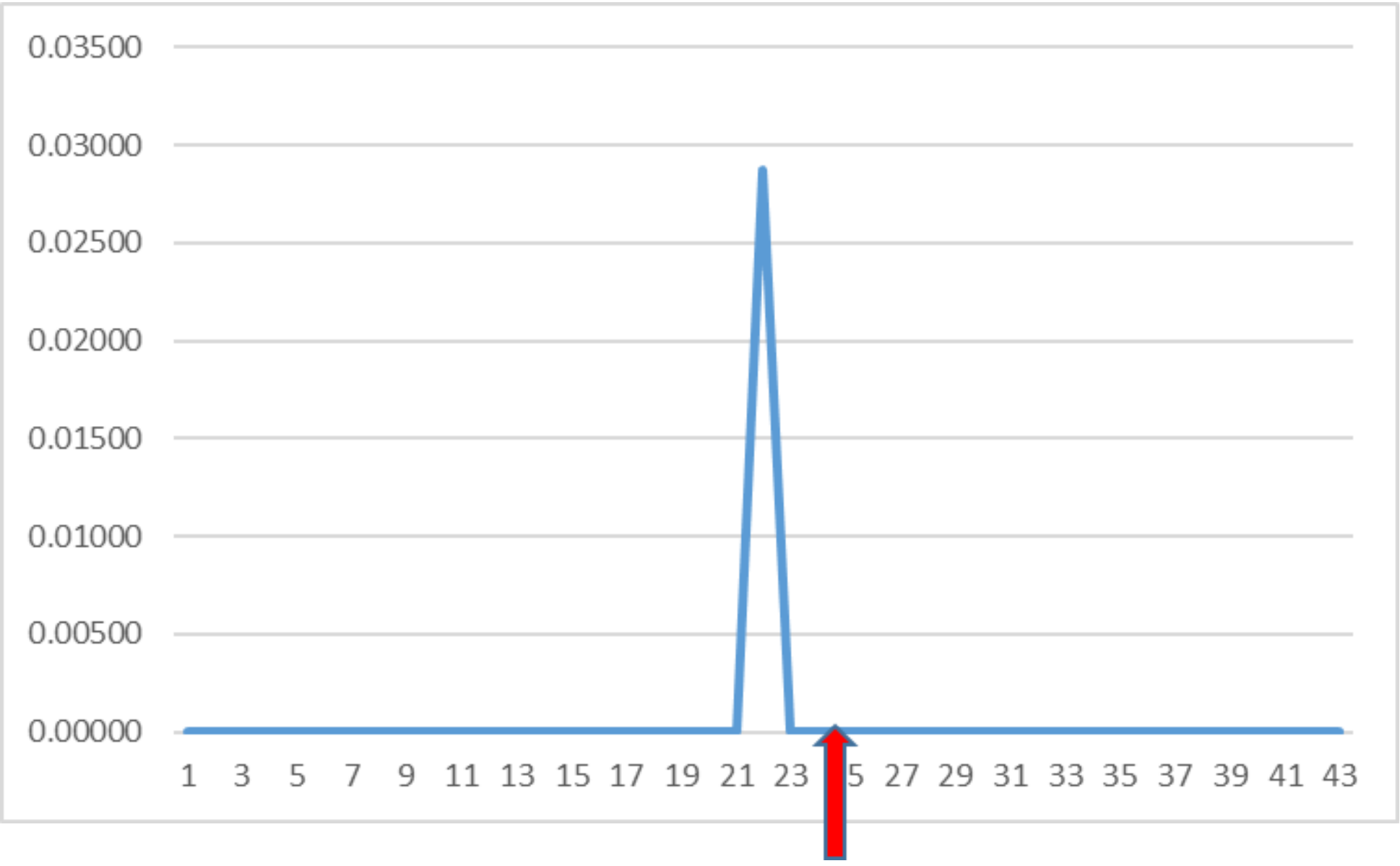

Fig.9. EM emissions “silence” period and anomaly, calculated by frequency ratio $\bar{x} \pm 3\sigma$ values

Research analysis of Crete earthquake conducted by **INFREP** data confirms the theoretical suggestions and experimental outcomes of scientific works published previously (Mjachkin et al.,1975; Freund et al., 2006; Eftaxias et al., 2009; Contadakis et al., 2010; Ouzounov et al., 2011; Biagi et al, 2013; Kachakhidze et al., 2015, 2016 (A)).

Methods offered in the presented article provide a large earthquake forecasting possibilities in case of monitoring of earthquake preparation process.

## 3. Conclusions

Based on the INFREP data, on example of Crete 25/05/2016- 08:36:13 UTC, M=5.6 earthquake, methods of large earthquake prediction have been developed, which in case of EM radiation monitoring, enable to determine epicenter, magnitude and time of occurring of incoming earthquake simultaneously.

In results of searching the following conclusions are made:

1) About 50 days prior to the earthquake, it is possible to separate a continuous active frequency channel;

2) By the active channel frequency, about 50 days before the earthquake, it is possible to determine the length of "cracked strip" on which the process of cracks origination is going on actively and ultimately the main fault forms.

3) By the length of the "cracked strip", it is possible to determine magnitude of icoming earthquake with certain accuracy about 50 days prior to the earthquake.

4) Once the active frequency channel detection, it already is possible to determine the future earthquake epicenter with certain accuracy;

5) In order to short-term prediction of a large earthquake, it is recommended to begin the frequency data monitoring from the starting moment of the avalanche-unstable process of fault formation and keep an eye on the process dynamics;

6) In the case of monitoring of electromagnetic emissions existent before earthquake, it is possible, step-by-step, to make about 50, 20 as well as 2-days short-term prediction of incoming earthquake;

7) EM emissions turned out to be the unique precursor, which is capable of the short-term prediction of the large earthquake.

REFERENCES

Biagi, P.F.: Seismic Effects on LF Radiowaves, in: Atmospheric and Ionospheric Electromagnetic Phenomena Associated with Earthquakes, (Ed.: M. Hayakawa), TERRAPUB, Tokyo, 1999, 535-542;

Biagi, P. F., Castellana, L., Maggipinto, T., Loiacono, D., Schiavulli, L., Ligonzo, T., Fiore, M., Suciu, E., and Ermini, A.: A pre seismic radio anomaly revealed in the area where the Abruzzo earthquake (M=6.3) occurred on 6 April 2009, Nat. Hazards Earth Syst. Sci., 9, 1551–1556, 2009, http://www.nat-hazards-earth-syst-sci.net/9/1551/2009/.

Biagi, P.F., Magippinto, T., Schiavulli, L., Ligonzo, T., Ermini, A.: European Network for collecting VLF/LF radio signals (D5.1a). DPC- INGV - S3 Project. "Short Term Earthquake prediction and preparation", 2013;

Bleier, T., Dunson, C., Maniscalco, M., Bryant, N., Bambery, R., and Freund, F.: Investigation of ULF magnetic pulsations, air conductivity changes, and infra red signatures associated with the 30 October Alum Rock M5.4 earthquake, Nat. Hazards Earth Syst. Sci., 9, 585–603, doi:10.5194/nhess-9-585-2009;

Boudjada, M.Y.; Schwingenschuh,K., Döller, R., Rohznoi, A., Parrot,M., Biagi, P. F., Galopeau, P.H.M., Solovieva, M., Molchanov,O., Biernat,H.K., Stangl, G, Lammer, H., Moldovan, I. Voller.W, and Ampferer, M. Decrease of VLF transmitter signal and Chorus-whistler waves before l'Aquila earthquake occurrence. Nat. Hazards Earth Syst. Sci., 10, 1487–1494, 2010. www.nat-hazards-earth-syst-sci.net/10/1487/2010/. doi:10.5194/nhess-10-1487-2010;

Contadakis, M. E., P. F. Biagi, and M. Hayakawa. Ground and satellite based observations during the time of the Abruzzo earthquake, Special Issue, Nat. Hazards Earth Syst. Sci., issue102, 2010;

Dobrovolsky, I.P., S.I. Zubkov and V.I. Miachkin. Estimation of the size of earthquake preparation zones, Pure Appl. Geophys., 117, 1025-1044, 1979;

Dudkin, F., Korepanov,V., Hayakawa. M., De Santis,A. Possible model of electromagnetic signals before earthquakes. "Thales", in honour of Prof. Emeritus M.E. Contadakis, Ziti Publishing, Thessaloniki, 2013;
http://www.topo.auth.gr/greek/ORG_DOMI/EMERITUS/TOMOS_CONTADAKIS/Contadakis_Honorary_Volume.htm

Eftaxias, K., L. Athanasopoulou, G. Balasis, M. Kalimeri, S. Nikolopoulos, Y. Contoyiannis, J. Kopanas, G. Antonopoulos and C. Nomicos. Unfolding the procedure of characterizing recorded ultralow frequency, kHZ and MHz electromagnetic anomalies prior to the L'Aquila earthquake as preseismic ones – Part 1, Nat. Hazards Earth Syst. Sci., 2009, 9, 1953– 1971;

Eftaxias, K., Potirakis,S.M., and Chelidze,T. On the puzzling feature of the silence of precursory electromagnetic emissions. Nat. Hazards Earth Syst. Sci., 2013,13, 2381–2397; www.nat-hazards-earth-syst-sci.net/13/2381/2013/. doi:10.5194/nhess-13-2381-2013;

Freund, F.T., Takeuchi, A., Lau, B.W.S., 2006. Electric currents streaming out of stressed igneous rocks – a step towards understanding pre-earthquake low frequency EM emissions. Phys. Chem. Earth 31, 389–396.

Hattori, K., Takahashi, I., Yoshino, C., Isezaki, N., Iwasaki, H., Harada, M., Kawabata, K., Kopytenko, E., Kopytenko, Y., Maltsev, P., Korepanov, V., Molchanov, O., Hayakawa, M., Noda, Y., Nagao, T., Uyeda, S. ULF geomagnetic field measurements in Japan and some recent results associated with Iwateken Nairiku Hokubu earthquake in 1998, Physics and Chemistry of the Earth, 29 (4-9) 481-494, 2004;

Hayakawa, M., Y. Hobara, A. Rozhnoi, M. Solovieva, K. Ohta, J. Izutsu, T. Akamura, Y. Yasuda, H. Yamaguchi and Y.Kasahara. The ionospheric precursor to the 2011 March 11 earthquake as based on the Japan-Pacific subionospheric VLF/LF network observation.“Thales”, in honour of Prof. Emeritus M.E. Contadakis, Ziti Publishing, Thessaloniki, 2013, pp. 191-212;.

Jenkins G. M. and Watts, D. G. “Spectral Analysis and Its Applications,” Mir, M, 1972. 287 p. (In Russian).

Johnston M. Review of electric and magnetic fields accompanying seismic and volcanic activity. Surveys in Geophysics 18: 441–475, 1997.

Kachakhidze, M.K., Kachakhidze,N.,K., Kaladze,T.D. A model of the generation of electromagnetic emissions detected prior to earthquakes. Physics and Chemistry of the Earth 85–86 2015, 78–81;

Kachakhidze Manana, Kachakhidze Nino, Kaladze Tamaz, Explanation of Litosphere-Atmosphere-Ionosphere Coupling System Anomalous Geophysical Phenomena on the Bases of the Model of Generation of Electromagnrtic Emissions Detected Before Earthquake. GESJ: Physics, No.2(14), pp. 66-75, 2015;

Kachakhidze M., Kachakhidze-Murphy N, Biagi P.F**.** Earthquake Forecasting Possible Methodology. GESJ: Physics. 2016 | No.1(15);

Kachakhidze M., Kachakhidze N. VLF/LF EM emissions as main precursor of earthquakes and their searching possibilities for Georgian s/a region. Geophysical Research Abstracts Vol. 18, EGU2016-3280, (A) 2016;

Mjachkin, V.I., W.F. Brace, G.A. Sobolev and J.H. Dieterich. Two models for earthquake forerunners, Pageoph., Basel. 1975, vol. 113;

Moldovan, I. A., A. O. Placinta, A. P. Constantin, Moldovan A. S. and C. Ionescu. Correlation of geomagnetic anomalies recorded at Muntele Rosu Seismic Observatory (Romania) with earthquake occurrence and solar magnetic storms, Ann. Geophys., 2012, 55,125–137, doi:10.4401/ag-5367;

Moroz, Y. F., T. A. Moroz, V. P.Nazareth, S. A. Nechaev and S.E. Smirnov. Electromagnetic field in studies of geodynamic processes. Complex seismological and geophysical researches of Kamchatka, (In Russian), 2004;

Ouzounov, Dimitar; Pulinets, Sergey; Romanov, Alexey; Romanov, Alexander; Tsybulya, Konstantin; Davidenko, Dmitri; Kafatos, Menas; Taylor, Patrick. Atmosphere-ionosphere response to the M 9 Tohoku earthquake revealed by multi-instrument space-borne and ground observations: Preliminary results. Earthquake Science, Volume 24, Issue 6, pp 557–564, 2011.

Ouzounov, Dimitar; Pulinets, Sergey; Hattori, Katsumi; Taylor, Patrick. Pre-Earthquake Processes: A Multidisciplinary Approach to Earthquake Prediction Studies.WILEY, 2018;

Papadopoulos, G. A., Charalampakis, M., Fokaefs, A., and Minadakis, G. Strong foreshock signal preceding the L'Aquila (Italy) earthquake (Mw6.3) of 6 April 2009, Nat. Hazards Earth Syst. Sci., 10, 19–24, doi:10.5194/nhess-10-19-2010;

Parrot, M. First results of the DEMETER micro-satellite, Planet. Space Sci., 2006, 54, 411-557;

Tramutoli, Valerio; Aliano, Carolina; Corrado, Rosita; Filizzola, Carolina; Genzano, Nicola; Lisi, Mariano; Martinelli, Giovanni; Pergola, Nicola. On the possible origin of thermal infrared radiation (TIR) anomalies in earthquake-prone areas observed using robust satellite techniques (RST). Elsevier, Chemical Geology, V. 339, p. 157-168, 2013

Varotsos, P., Sarlis, N., Skordas, E., and Lazaridou, M.: Additional evidence on some relationship between Seismic Electric Signals (SES) and earthquake focal mechanism, Tectonophysics, 412, 279–288, 2006;